# Finite-word-length FPGA implementation of model predictive control for ITER resistive wall mode control

Samo Gerkšič[a*], Boštjan Pregelj[a]

[a]*Jožef Stefan Institute, Ljubljana, Slovenia*

Abstract: In advanced tokamak scenarios, active feedback control of unstable resistive wall modes (RWM) may be required. A RWM is an instability due to plasma kink at higher plasma pressure, moderated by the presence of a resistive wall surrounding the plasma. We address the dominant kink instability associated with the main non-axisymmetric (n = 1) RWM, described by the CarMa model. Model predictive control (MPC) is used, with the aim of enlarging the domain of attraction of the unstable RWM modes subject to power-supply voltage constraints. The implementation of MPC is challenging, because the related quadratic programming (QP) on-line optimization problems must be solved at a sub-ms sampling rate. Using complexity-reduction pre-processing techniques and a primal fast gradient method (FGM) QP solver, sufficiently short computation times for ITER are reachable using a standard personal computer (PC). In this work we explore even faster finite-word-length (FWL) implementation using field-programmable gate arrays (FPGA), which would facilitate experimental testing of such control algorithms on dynamically faster medium-sized tokamaks, and compare the computational accuracy and time with the PC implementation.

## 1. Introduction

Tokamak control systems must deal with different kinds of plasma instabilities. The most common is the vertical instability due to an axisymmetric (n = 0) mode. A resistive wall mode (RWM) is an instability due to plasma kink at higher normalized plasma pressures $\beta_N$, moderated by the presence of a resistive wall surrounding the plasma [1, 2, 3]. Passive RWM stabilization approaches based on plasma rotation and kinetic effects may not be sufficient [4, 5]. The subject of this work is active feedback stabilization of the kink instability associated with the main non-axisymmetric (*n* = 1) RWM, according to the modelling expected to be dominant in ITER advanced tokamak scenarios.

RWM control has been established in experiments on reversed-field pinch devices RFX-mod [6], EXTRAP T2R [7] and tokamaks HBT-EP [8], DIII-D [9], and NSTX [10]. RWM modelling codes, such as VALEN [11] and CarMa [12] are based on coupling the plasma MHD equations with the eddy current equations for the conductive wall. RWM control experiments were mostly performed with easily reproducible current-driven RWMs [6], or by simulating RWMs with a different set of control coils [13]. Pressure-driven RWMs [10] are more relevant for advanced tokamak scenarios, but they are less reproducible and are often accompanied by other types of MHD instabilities, including edge localized modes (ELM), neoclassical tearing modes (NTM), etc. [2, 8]. Magnetic active feedback RWM control is closely related to error field correction (EFC) [9]. EFC is intended for the compensation of the static error field due to the geometrical imperfections and of the resonant perturbation of the plasma to the static error field. Both EFC and active RWM control act on correction coils, located either internally or externally to the vacuum vessel. They may even use the same coil set [9, 10], in some cases within the same controller [7]. However, they act in different frequency bands – EFC compensates the static error field and the low-frequency time-variation of the plasma response to it, while active RWM feedback operates in the higher-frequency range. Internal coils are preferred for active RWM control because of the conductive wall shielding the external coils. In ITER, three sets of six external superconducting correction coils are planned for EFC [14], while active RWM feedback may be implemented using three sets of nine internal ELM coils [15] as a secondary function. In this work, it is assumed that a separate EFC system will be used.

The initial RWM feedback control approaches were based on multiple local single-input single-output (SISO) proportional (P) or proportional-derivative (PD) control of actuator coils based on adjacent sensor measurements of the radial and/or poloidal magnetic field [9, 4]. Later, several advanced model-based control approaches have been applied, including linear quadratic Gaussian (LQG) optimal control [11, 10, 7, 15, 16, 13] and model predictive control (MPC) [17, 18]. The control-oriented model required for these approaches can be obtained either with experimental or with first-principles modelling and model simplification/reduction procedures. The advanced control approaches are able to make better use of the available control coils [13] and enlarge the stabilizable region of the unstable RWM modes subject to actuator voltage constraints [16].

MPC is an advanced process control method that has become well established in the process industry. It provides a systematic approach to control large-scale multivariable systems and efficient handling of constraints on process variables. These advantages are beneficial to tokamak control, but the relatively long sample computation time typically

---

\* *Corresponding author. Email: Samo.Gerksic@ijs.si, postal address: IJS E2, Jamova 39, 1000 Ljubljana, Slovenia*

needed for solving the on-line optimization problem, typically in the form of a QP, presents a difficult obstacle. Accelerating MPC has been a topic of intense research recently [19, 20], including in the nuclear fusion control community [17, 21, 23, 24].

RWM control with MPC is challenging because very fast dynamics are combined with high system dimensions (27 actuators, 6 sensors, 50 dynamic states). Therefore, a pre-computed (multiparametric) MPC approach [25] is not applicable, and the dual Fast Gradient Method (FGM) solver [20, 21] exceeds the allowed computation time, which should be less than 0.1 ms. Our recent work [26] shows that an infinite-horizon MPC considering only actuator constraints using a solver such as the primal FGM is viable. The MPC RWM controller is related to the LQG RWM scheme for ITER of Ariola and Pironti [16], based on the CarMa model [12], however without control vector dimension reduction. The MPC controller also uses the Kalman filter (KF) [11] for system state vector estimation from system output measurements. In simulations, the MPC controller stabilises the system with narrower voltage constraints and using less power than the LQG controller and with a computation time 0.1 ms that is acceptable for RWM control in ITER, using a standard desktop PC. Differences between LQG and MPC control are most notable in cases of RWM perturbations with large amplitudes. With the linear LQG controller, when a large actuator voltage signal exceeding the saturation limit is demanded, merely saturation is applied. The MPC controller is aware of the approaching saturation limit, and is therefore able to increase the actuator voltage in advance to generate the suppressive magnetic field needed to stabilize the RWM disturbance. Further, the MPC controller is in such case able to employ other relevant un-saturated actuators at the same time, while the LQG controller does so with a delayed feedback action. However, such control algorithms need to be tested experimentally on smaller tokamak devices with faster dynamics, which require much shorter sampling times.

Due to the iterative nature of the QP solver which allows only micro-parallelization of matrix-vector arithmetic operation within iterations and very short computation times, speeding up the computation substantially by multi-threading on standard multi-core central processing units is not straightforward. It is known that the desired acceleration can be achieved by FPGA implementation [19, 17] using hardware description language (HDL) programming; however, such low-level programming requires specific FPGA programming skills, and is known to take a lot of time for implementation. Recently, high-level synthesis (HLS) approaches to FPGA programming are gaining popularity, because they facilitate faster FPGA implementation of algorithms without in-depth HDL programming experience and avoiding manual coding errors in the translation process [27]. In a recent attempt [28], the Xilinx Vivado HLS environment was used to convert the dual Fast Gradient Method (FGM) QP solver used for MPC control for ITER plasma current and shape control from C code to HDL code suitable for implementation on a Xilinx ZC706 FPGA board. Moderate performance improvement without excessive use of resources was achieved by using emulated single-precision floating-point arithmetic. More substantial acceleration would be possible by using a larger FPGA or an FPGA with hardware floating-point computational units (e.g., Altera Arria 10), both also at substantially higher equipment cost.

Computational implementation of algorithms using finite word-length (FWL) arithmetic tends to be more efficient for FPGA programming, both regarding the speed of computation and the use of FPGA resources [29, 30]. In particular, the Xilinx Vivado HLS environment supports "arbitrary-precision fixed-point" (APF) number representation, however it does not provide much aid in conversion from an original floating-point algorithm, which is not a trivial step with algorithms of such scale. To this end, we found Matlab Fixed-Point Designer to be a practical tool for selecting value ranges and bit-widths needed for the chosen computation accuracy. It would be convenient to follow up with conversion of Matlab code to HDL code using Matlab HDL Coder, but its workflow from Matlab code has many technical limitations and does not allow efficient use of FPGA resources; HDL Coder workflow from Simulink appears more elaborate, but would require extensive manual recoding of the algorithm into graphical form using a specific block set. Hence, for the conversion to HDL we choose the workflow from C++ using Xilinx Vivado HLS 2019.1, also involving a manual restructuring of the code and adding compiler directives. The hardware design target was the upper-midrange Xilinx Kintex UltraScale+ KCU116 evaluation board. Due to equipment availability, the actual testing was performed with a higher-tier Xilinx Alveo U250 accelerator card, using the Xilinx Vitis 2019.1 environment.

In this work we describe an infinite-horizon MPC controller for ITER RWM employing a primal FGM QP solver and its conversion from floating-point Matlab m-code to FWL HDL FPGA implementation. The conversion process involved the Matlab Fixed-Point Converter for FWL conversion, then Xilinx Vivado HLS as the HLS environment, and Xilinx Vitis for the final implementation. The outcome of the conversion process is compared to the C-code primal FGM QP solver on a standard desktop PC in terms of the numerical accuracy and the execution time. In Section 2 we describe the ITER RWM control problem setup. Section 3 reviews the design of the proposed infinite-horizon MPC controller with the KF with scaled input, output and state vectors. Section 4 outlines the primal FGM QP solver. In Section 5, the FWL conversion procedure and code modification are described. Section 6 presents the performance evaluation results.

## 2. Overview of RWM control for ITER

The control problem setup [26] is based on that of Ariola and Pironti [16]. In Figure 1, the resistive in-vessel non-axi-symmetric ELM correction coils that are used as actuators for RWM suppression are drawn with light blue colour. The

27 ELM coils are arranged in three sectors vertically and nine sectors horizontally along the toroidal vessel; the horizontal sectors are located at the toroidal angles $\varphi_{ELM, i} = 40°·(i-1)$, with $i = 1, ..., 9$. The vector of coil power-supply (PS) voltages applied to the ELM coils is $\mathbf{u}_{ELM} \in \Re^{27}$, and $\mathbf{u} \in \Re^{27}$ is the control signal vector at the inputs of the PS. For our MPC control, the reduction of $\mathbf{u}$ to 6 elements as in [16] is not used, because simple bounds on the actuator signal are preferred for the primal FGM solver.

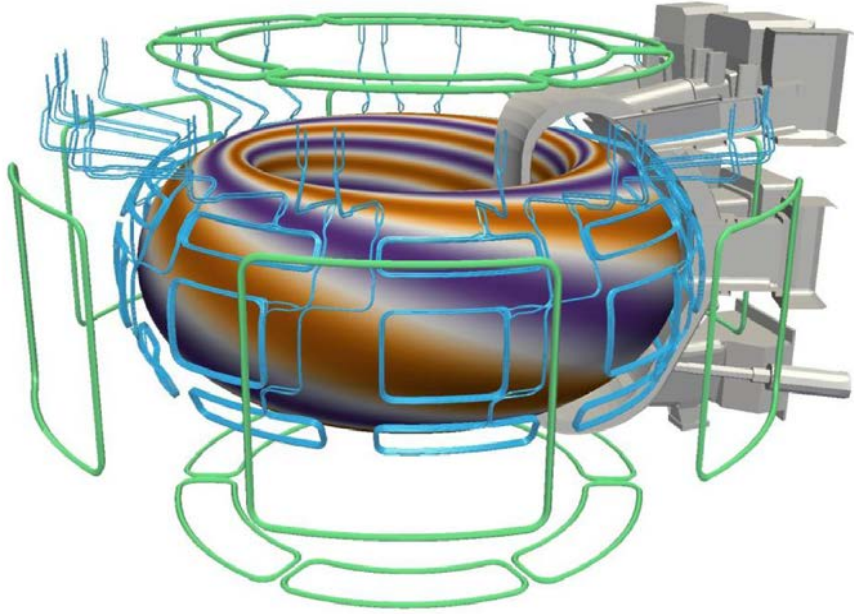

Figure 1. ITER non-axi-symmetric correction coils: in-vessel ohmic ELM coils (light blue) and external superconductive EFC coils (light green). Image © ITER.org, used with permission.

For control design, the power supplies are modelled as a serial interconnection of saturation block with the limits $[V_{min}, V_{max}] = [-144$ V, $144$ V$]$, a first-order lag transfer function with the time constant $7.5 \cdot 10^{-3}$, and a $2.5 \cdot 10^{-3}$ $s$ time delay block approximated with a 3-rd order Padé function ([31], section 5.2). The absolute value of the ELM coil currents should not exceed $1.5 \cdot 10^4$ A. The design limits of voltages and currents of the ELM coils are much higher, however RWM suppression is their secondary function.

For RWM detection, an array of 6 vertical magnetic field $B_v$ sensors located at the poloidal position ($r = 8.928$ m, $z = 0.550$ m), toroidally spaced at 60° angles $\varphi_i \in \{39°, 101°, 159°, 221°, 279°, 341°\}$ is used. Their measurement values comprise the measurement output signal $\mathbf{y}_m \in \Re^6$. The control scheme is intended for suppression the of $n = 1$ mode that may be described with its cosine and sine components as

$$y_i = y_A \cos\varphi_i + y_B \sin\varphi_i, \quad i \in \{1, 2, ..., 6\} \quad (1)$$

The reduced-dimensional system output vector $\mathbf{y} \in \Re^2 = [y_A\ y_B]^T$ [16] is determined via the least squares method

$$\mathbf{y} = \mathbf{T}_{out}\mathbf{y}_m, \quad \mathbf{T}_{out} = \begin{bmatrix} \cos\varphi_1 & \sin\varphi_1 \\ \cos\varphi_2 & \sin\varphi_2 \\ \vdots & \vdots \\ \cos\varphi_6 & \sin\varphi_6 \end{bmatrix}^\dagger \quad (2)$$

The RWM simulation model is generated by the CarMa code [12], which combines three-dimensional axi-symmetric linearised MHD equations for the plasma, and equations for the currents induced in the conductors of the vessel wall, discretized to a mesh of hexahedral elements. The following linear state-space model describing perturbations from the equilibrium configuration for the selected ITER plasma equilibrium is obtained (all values are displacements from the equilibrium values, although the difference operator Δ is omitted)

$$\dot{\mathbf{x}}_{CM} = \mathbf{A}_{CM}\mathbf{x}_{CM} + \mathbf{B}_{CM}\mathbf{u}_{ELM} \quad (3)$$

$$\mathbf{y}_m = \mathbf{C}_{CM,m}\mathbf{x}_{CM} \quad (4)$$

where the state vector $\mathbf{x}_{CM} \in \Re^{6294}$ includes the current displacements, and $\mathbf{A}_{CM}$, $\mathbf{B}_{CM}$, $\mathbf{C}_{CM,m}$ are the RWM dynamics state-space model matrices, respectively. $\mathbf{A}_{CM}$ has two unstable eigenvalues corresponding to the cosine and sine components of the unstable RWM mode, with the growth rate $\gamma$ (about 19 s$^{-1}$) and frequency $\omega$ (about 0.26 s$^{-1}$). The model also includes an auxiliary output vector of ELM coil currents $\mathbf{I}_{ELM} \in \Re^{27}$ via the auxiliary output matrix $\mathbf{C}_{CM,ELM}$

$$\mathbf{I}_{ELM} = \mathbf{C}_{CM,ELM}\mathbf{x}_{CM} \tag{5}$$

## 3. Model Predictive Control

The main objective of using MPC for RWM control is improving the control performance in the areas close to the actuator coil voltage constraints, facilitating the enlargement of the domain of attraction in which the unstable RWM modes can be stabilized. MPC design comprises building the control-oriented model, specifying the infinite-horizon MPC cost function and constraints, and designing the Kalman filter (KF). Certain design choices are motivated by the requirement for very fast implementation with the primal FGM QP solver [26].

The control-oriented model of the controlled system, which includes the PS dynamics and the RWM dynamics (3-4), in a suitable discrete-time state-space form is needed

$$\mathbf{x}(k + 1) = \mathbf{A}\mathbf{x}(k) + \mathbf{B}\mathbf{u}(k) \tag{6}$$
$$\mathbf{y}(k) = \mathbf{C}\mathbf{x}(k) \tag{7}$$

where $k$ is the discrete time index, $\mathbf{A}$, $\mathbf{B}$, and $\mathbf{C}$ are the system matrices obtained via zero-order-hold discretisation with the sampling time $T_s = 0.75$ ms, and $\mathbf{x} \in \Re^{N_x}$ is the state vector, with the model order $N_x = 50$ reduced as much as possible while still covering the relevant frequency range of dynamics.

The infinite-horizon MPC controller cost function consists of two parts: a finite-horizon MPC cost for the first part of the future horizon from sample $k$ up to sample $(k + N - 1)$ within which the control signal constraints are considered, and a terminal infinite-horizon LQ cost for samples from $N$ onwards, where $N$ denotes the length of the finite horizon. The cost function is

$$J(k) = \tfrac{1}{2}\sum_{i=0}^{N-1}\left(\mathbf{x}_{k+i}^T\mathbf{Q}_C\mathbf{x}_{k+i} + \mathbf{u}_{k+i}^T\mathbf{R}_C\mathbf{u}_{k+i}\right) + \tfrac{1}{2}\mathbf{x}_{k+N}^T\mathbf{P}\mathbf{x}_{k+N} \tag{8}$$

where the lower discrete-time indices denote signal prediction based on the current state $\mathbf{x}(k)$; $\mathbf{Q}_C$ and $\mathbf{R}_C$ are state and control cost matrices, respectively; the term $\mathbf{x}_{k+N}^T\mathbf{P}\mathbf{x}_{k+N}$ represents the terminal LQ cost, and $\mathbf{P}$ is the steady-state solution of the discrete-time algebraic Riccati equation [11].

The MPC control problem to be solved with QP optimization at each sample time is

$$\min_{\mathbf{u}_k,\ldots,\mathbf{u}_{k+N-1}} J(k)$$

$$\text{subject to } \mathbf{x}_{k+j+1} = \mathbf{A}\mathbf{x}_{k+j} + \mathbf{B}\mathbf{u}_{k+j}, \quad j = 0, 1, \ldots, N - 1$$

$$\mathbf{u}_{\min} \leq \mathbf{u}_{k+j} \leq \mathbf{u}_{\max}, \quad j = 0, 1, \ldots, N - 1$$

$$\mathbf{x}_k = \mathbf{x}(k) \tag{9}$$

The steady-state Kalman filter (KF) is used to determine the current state estimate $\mathbf{x}(k|k)$, which is used in place of the non-measurable system state $\mathbf{x}(k)$ [11]

$$\mathbf{x}(k|k-1) = \mathbf{A}\mathbf{x}(k-1|k-1)\mathbf{x}(k-1|k-1) + \mathbf{B}\mathbf{u}(k-1) \tag{10}$$
$$\mathbf{x}(k|k) = \mathbf{x}(k|k-1) + \mathbf{M}_K[\mathbf{y}(k) - \mathbf{C}\mathbf{x}(k|k-1)] \tag{11}$$

where $\mathbf{M}_K$ is computed via the steady-state solution of the Riccati equation from the covariance matrices $\mathbf{Q}_K = E\{\mathbf{w}\mathbf{w}^T\}$ and $\mathbf{R}_K = E\{\mathbf{v}\mathbf{v}^T\}$, where $\mathbf{w} \in \Re^{N_x}$ and $\mathbf{v} \in \Re^2$ are the assumed white noise disturbance vector signals entering at the system state and output, respectively. However, we use the KF as an observer, so that the diagonal elements of $\mathbf{Q}_K$ and $\mathbf{R}_K$ are used as tuning parameters to achieve desired dynamics.

For the FWL implemementation of the controller using Matlab Fixed-Point Designer, scaling of the $\mathbf{u}$, $\mathbf{x}$, and $\mathbf{y}$ vectors is required, because separate value ranges for individual vector elements are not supported. Vector scaling is introduced

$$\mathbf{u} = \mathbf{K}_\mathbf{u}\mathbf{u}_s, \quad \mathbf{x} = \mathbf{K}_\mathbf{x}\mathbf{x}_s, \quad \mathbf{y}_s = \mathbf{K}_\mathbf{y}\mathbf{y} \tag{12}$$

where the scaled vectors, denoted with the lower index s, have all elements in ranges [−1, 1], and $\mathbf{K}_\mathbf{u}$, $\mathbf{K}_\mathbf{x}$ and $\mathbf{K}_\mathbf{y}$ are diagonal scaling matrices of appropriate dimension. The scaled state-space model matrices are

$$\mathbf{A}_s = \mathbf{K}_\mathbf{x}^{-1}\mathbf{A}\mathbf{K}_\mathbf{x}, \qquad \mathbf{B}_s = \mathbf{K}_\mathbf{x}^{-1}\mathbf{B}\mathbf{K}_\mathbf{u},$$
$$\mathbf{C}_s = \mathbf{K}_\mathbf{y}\mathbf{C}\mathbf{K}_\mathbf{x}, \qquad \mathbf{D}_s = \mathbf{K}_\mathbf{y}\mathbf{D}\mathbf{K}_\mathbf{u} \tag{13}$$

The tuning matrices are also scaled, so that the same tuning as without scaling is achieved

$$\mathbf{Q}_{Cs} = \mathbf{K_x}\,\mathbf{Q}_C\mathbf{K_x}, \qquad \mathbf{R}_{Cs} = \mathbf{K_u}\,\mathbf{R}_C\mathbf{K_u},$$

$$\mathbf{Q}_{Ks} = \mathbf{K_x^{-1}}\mathbf{Q}_K\mathbf{K_x^{-1}}, \qquad \mathbf{R}_{Ks} = \mathbf{K_y}\,\mathbf{R}_{Ks}\mathbf{K_y} \tag{14}$$

so that the controller design is carried out using the scaled model and tuning matrices in place of the original ones. In the Matlab/Simulink control scheme in Figure 2, signal scaling blocks are inserted at the controller output and at the KF inputs. The MPC problem (9) of the MPC controller block may be solved by the primal FGM solver derived from the QPgen toolbox [20] in the floating-point or FWL implementation, or for reference using ILOG/IBM CPLEX as the QP solver. The original high-order model of plasma RWM dynamics (3-4) is used as the plasma model. The actuator voltage constraints are set to $|\mathbf{u}| \leq 34$ V.

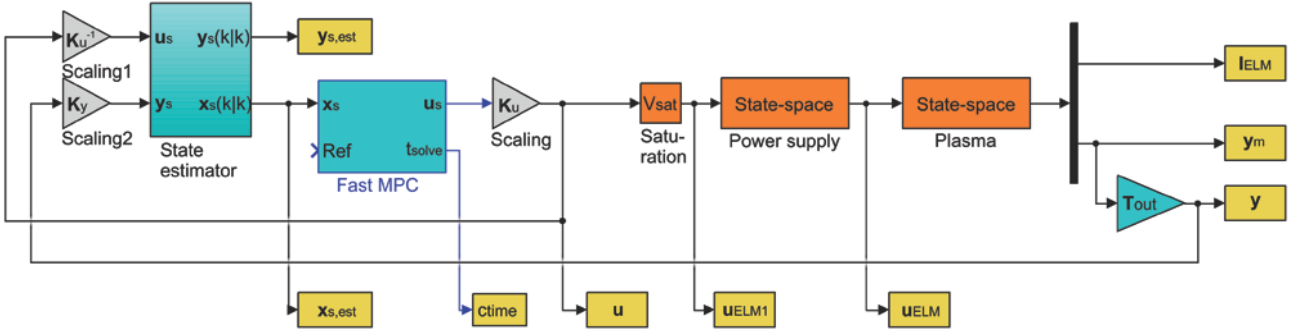

Figure 2. Simulink block diagram for MPC ITER RWM control with scaling

The controller tuning is a compromise between the requirements for controller responsiveness needed for the stabilisation of the unstable modes and the suppression of measurement noise. Figure 3 displays simulation of system stabilization from a perturbed initial state with the MPC controller with the horizon $N = 80$ and move blocking to 3 intervals of lengths (2, 2, 76). Move blocking is very efficient for reduction of the computational load at longer horizons. Typically, we use blocking of the control signal within the predictive horizon to $N_u = 3$ intervals of lengths (2, 2, 76), resulting in additional equality constraints $\mathbf{u}_0 = \mathbf{u}_1$, $\mathbf{u}_2 = \mathbf{u}_3$, and $\mathbf{u}_4 = \mathbf{u}_5 = ... = \mathbf{u}_{N-1}$.

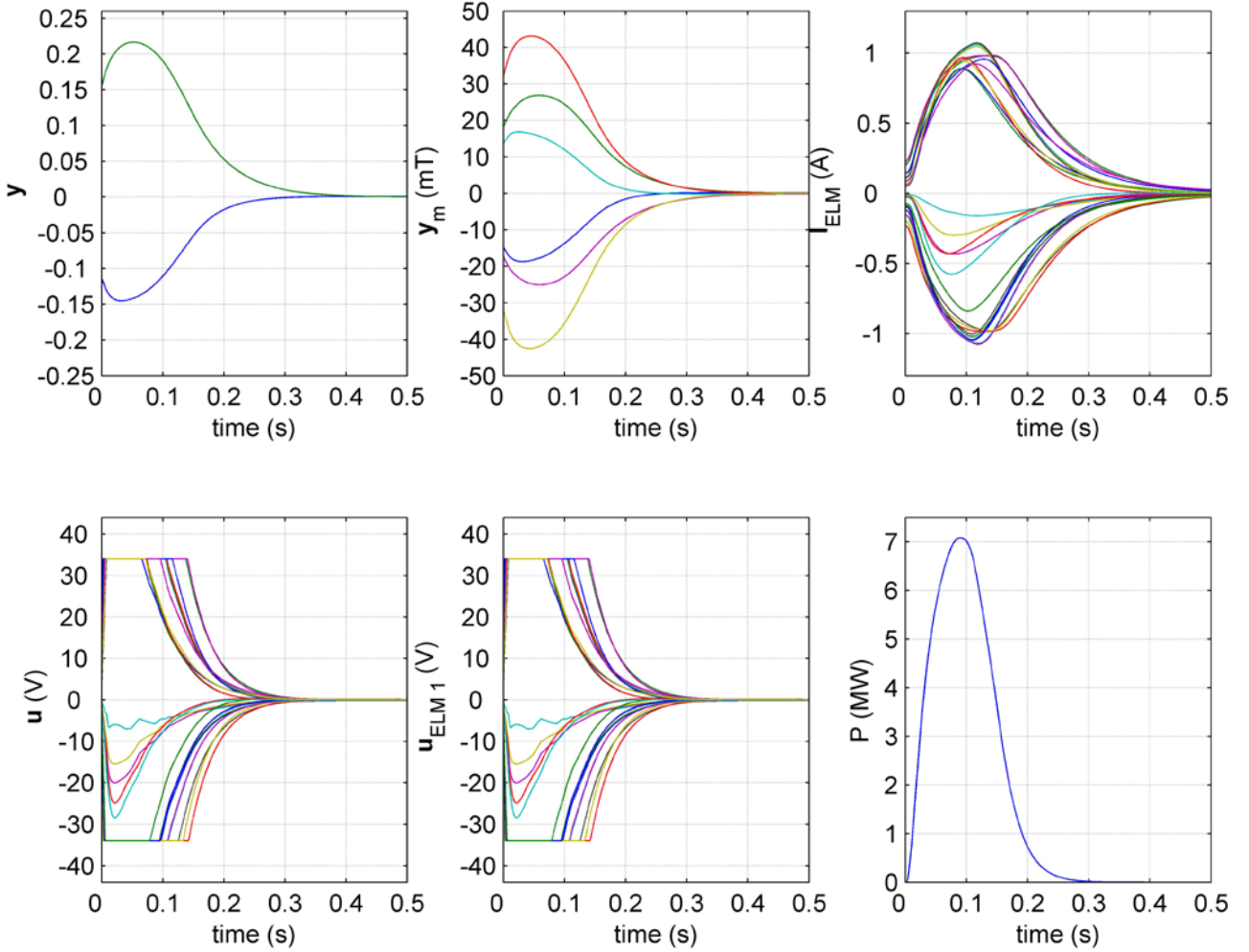

Figure 3. MPC control simulation with signal scaling. Signals: reduced-dimensional measurement vector **y** (top-left); measurement output signal $\mathbf{y}_m$ (top-centre); vector of ELM coil currents $\mathbf{I}_{ELM}$ (top-right); control signal vector (PS input) **u** (bottom-left); vector of PS voltages after saturation $\mathbf{u}_{ELM1}$ (bottom-centre); power at ELM coils (bottom-right).

## 4. Primal FGM QP solver

The QP solver is based on the generalized primal FGM algorithm [32] of the QPgen code generator [20]. This is a relatively simple optimization algorithm, known to be efficient for MPC control problems with input constraints. The convergence rate is not distinguished, but FGM iterations are simple, and not many are required typically for the precision demanded for control, and a practically meaningful relatively tight theoretical upper bound on the convergence range exists. FGM belongs to the family of first-order proximal optimization methods, designed for problems of the minimization of a sum of two functions with specific properties. In our MPC case [26], we are solving the problem

$$\min(J(\tilde{\mathbf{u}}) + \psi(\tilde{\mathbf{u}})) \qquad (15)$$

where the optimization variable is the vector of the controller outputs over the predictive horizon, $\tilde{\mathbf{u}} = [\mathbf{u}_k, \ldots, \mathbf{u}_{k+N-1}]$ (using the *condensed* approach, the system state **x** does not appear in the optimization variable, because the equality constraints of the control model state transition equation (6) are directly substituted into the MPC cost function (8); using move blocking, the dimension of $\tilde{\mathbf{u}}$ is reduced to $27 \cdot N_u$; the first function $J$ is the MPC cost function (8); the second function $\psi$ is the indicator function of the feasible set of the input inequality constraints

$$\psi(\tilde{\mathbf{u}}) = \begin{cases} 0 & \text{if } \tilde{\mathbf{u}}_{\min} \leq \tilde{\mathbf{u}} \leq \tilde{\mathbf{u}}_{\max} \\ \infty & \text{otherwise} \end{cases} \qquad (16)$$

The generalized proximity operator is defined as [20]

$$\operatorname{prox}_\psi^{\mathbf{L}}(\boldsymbol{\chi}) \triangleq \arg\min_{\mathbf{y}} \left( \psi(\mathbf{y}) + \frac{1}{2} \|\mathbf{y} - \boldsymbol{\chi}\|_{\mathbf{L}}^2 \right) \qquad (17)$$

and with the choice of the indicator function $\psi$ represents a simple projection onto a positive orthant, in practice implemented by clipping each vector element with its respective constraint value, and **L** is a diagonal preconditioning matrix.

The MPC QP being solved in the condensed form is formulated as [26]

$$\min_{\widetilde{\mathbf{u}}} \frac{1}{2} \widetilde{\mathbf{u}}^T \mathbf{H}_c \widetilde{\mathbf{u}} + \mathbf{f}_c^T \widetilde{\mathbf{u}} + c_c$$

$$\text{subject to } \begin{bmatrix} \mathbf{I} \\ -\mathbf{I} \end{bmatrix} \widetilde{\mathbf{u}} \leq \begin{bmatrix} \widetilde{\mathbf{u}}_{\max} \\ \widetilde{\mathbf{u}}_{\min} \end{bmatrix}, \quad (18)$$

where $\mathbf{H}_c$, $\mathbf{f}_c$, and $c_c$ are the second-order, first-order and constant cost function terms, respectively, and $c_c$ may be omitted as it does not affect the optimization result.

The FGM algorithm is summarized in Algorithm 1 below, where the top indices represent the iterations. It requires a specific scalar sequence $\beta^i$, which may be computed in advance, to achieve the desired convergence properties.

Algorithm 1: Fast gradient method
Initialize: $\mathbf{v}^1 = \widetilde{\mathbf{u}}^0 \in \mathbb{R}^{27N_u}$, sequence $\beta^i$, vector $\mathbf{f}_c$
**for** $i = 1$ to $i_{\max}$
    $\boldsymbol{\chi}^i = \mathbf{v}^i - \mathbf{L}^{-1}(\mathbf{H}_c \mathbf{v}^i + \mathbf{f}_c)$     (gradient step)
    $\widetilde{\mathbf{u}}^i = \text{prox}_\psi^{\mathbf{L}}(\boldsymbol{\chi}^i)$     (projection onto feasible set)
    $\mathbf{v}^{i+1} = \widetilde{\mathbf{u}}^i + \beta^i(\widetilde{\mathbf{u}}^i - \widetilde{\mathbf{u}}^{i-1})$     (acceleration)
    **if** $(\mathbf{v}^i - \widetilde{\mathbf{u}}^i)^T(\widetilde{\mathbf{u}}^i - \widetilde{\mathbf{u}}^{i-1}) > 0$     (adaptive restart)
        $\mathbf{v}^{i+1} = \widetilde{\mathbf{u}}^{i-1}, \quad \widetilde{\mathbf{u}}^i = \widetilde{\mathbf{u}}^{i-1}$
    **endif**
**endfor**

$\mathbf{L}$ is determined by solving an optimization problem seeking the optimal preconditioner with diagonal structure, which minimises the condition number of the scaled Hessian $\mathbf{H}_{cp} = \mathbf{L}^{-1}\mathbf{H}_c$ and sets the highest eigenvalue to 1 with the aim of improving the convergence [20]. In the practical implementation of Algorithm 1, the scaled QP is used, so that the multiplication with $\mathbf{L}^{-1}$ is not required in each iteration. Cold-starting from $\widetilde{\mathbf{u}}^0 = \mathbf{0}$ is used.

For assessing the accuracy of the FGM QP solver solution $\widetilde{\mathbf{u}}^i$, we compare it to the optimal solution $\widetilde{\mathbf{u}}^*$ (in practice, a reference solution computed using CPLEX). For this comparison we use the normalised mean-square-error (MSE) formula

$$MSE(\widetilde{\mathbf{u}}^i, \widetilde{\mathbf{u}}^*) = \sqrt{\frac{1}{27N_u} \sum_{k=1}^{27N_u} \left( \frac{\widetilde{u}_k^i - \widetilde{u}_k^*}{\widetilde{u}_{\max,k} - \widetilde{u}_{\min,k}} \right)^2} \quad (19)$$

A sufficiently low worst-case MSE below $10^{-4}$ among all samples of the simulation is achieved with $i = 20$ iterations of the FGM algorithm, at cost function deviation $J(\widetilde{\mathbf{u}}^i) - J(\widetilde{\mathbf{u}}^*)$ below $3 \cdot 10^{-4}$. According to the linear bound of the convergence rate [32, 22] it would require 50 iterations for a certified solution. Figure 4 shows convergence of the cost function $J$ with iterations of the algorithm for 12 example initial states and the two theoretical convergence bounds. Both theoretical and practical convergence are substantially slower if preconditioning is not used.

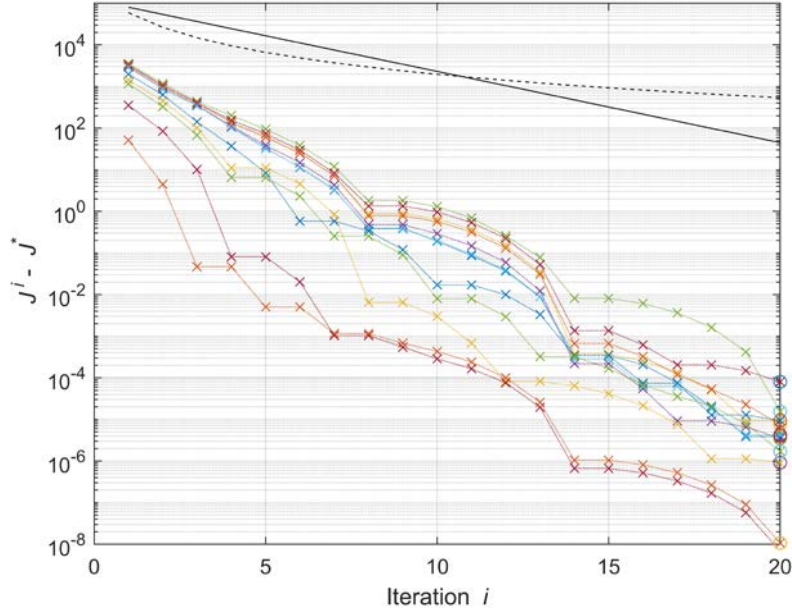

Figure 4. Primal FGM QP solver: convergence of the cost function value $J^i$ after $i$ iterations towards the optimal value $J^*$. 12 lines marked with '×' represent convergence sequences from different initial state estimates. The black solid and dashed line show the two theoretical upper convergence bounds.

## 5. Conversion to FWL FPGA implementation

The conversion to the FWL form was carried out with the aid of Matlab Fixed-Point Converter (FPC). FPC facilitates automated conversion of ordinary floating-point Matlab code to an adjustable-precision fixed-point FWL 'fi' format or to single-precision floating point. Despite certain restrictions on the Matlab language syntax, no particular code adaptation was needed for our implementation of the primal FGM QP solver function in Algorithm 1. One must prepare a "test-bench" function, which loads a relevant set of test input data, runs the FWL-converted function with all inputs in this set, and compares the FWL results with reference results in terms of the cost function $J$ and the MSE. FPC automatically records the value ranges of all variables for the data in the test set and suggests value ranges and bit-depths accordingly; but in order to avoid overflows, one should set known value ranges for the input value ranges, and use theoretical rules of value range and bit-depth propagation with arithmetic operations for derived variables. For the FGM algorithm, theoretical FWL implementation discussion is available in [19, 29, 30].

While the initial FWL implementation was easy to reach, the bit-widths required for the desired precision were excessive, particularly because the elements of the state vector **x** were of different orders of magnitude, and FPC assumes the same value ranges for all elements of a vector or a matrix. As described above, scaling to the value interval [−1, 1] according to Eq. (12) was introduced for the **u**, **x**, and **y** vectors. Even with this scaling, the algorithm implemented following QPgen [20] was still not best suited. As mentioned above, QPgen uses a modified implementation of Algorithm 1 in order to reduce the number of multiplication operations in the gradient step. In particular, the multiplication with $\mathbf{L}^{-1}$ of the gradient step in each iteration is avoided by working with variants of the vectors $\mathbf{v}^i$, $\mathbf{\chi}^i$, and $\tilde{\mathbf{u}}^i$ scaled with $\mathbf{L}^{-1}$ throughout the iterations, and all matrices and vectors for Algorithm 1 are accordingly prepared in the pre-processing phase. While this is appropriate for the floating-point implementation, it is not convenient with a fixed-point FWL implementation. The element of the diagonal preconditioner $\mathbf{L}^{-1}$ span about two orders of magnitude, hence about 6 more bits are required for the bit-width of the scaled variants of the vectors $\mathbf{v}^i$, $\mathbf{\chi}^i$, and $\tilde{\mathbf{u}}^i$ for the same precision of the computation. Therefore, we employ a variant of Algorithm 1 with a different scaling approach, where the vectors $\mathbf{v}^i$, $\mathbf{\chi}^i$, and $\tilde{\mathbf{u}}^i$ are not scaled, while the multiplication with $\mathbf{L}^{-1}$ in the gradient step of each iteration is still not needed. Instead, multiplication with $\mathbf{H}_{cp}$ is carried out in the gradient step, and other matrices and vectors for Algorithm 1 are suitably prepared in the pre-processing phase. Following these modifications, using bit-width 26 for most variables (64 for the restart test condition computation) resulted in the degradation of the computation accuracy due to FWL implementation below $8 \cdot 10^{-5}$ in the MSE and below $2 \cdot 10^{-6}$ in the cost function $J$.

### 5.1. C++ code adaptation

A substantial manual adaptation of the original C++ code of Algorithm 1 was required for the HLS conversion to HDL code using Xilinx Vivado HLS 2019.1.

Basically, the C++ language syntax is restricted, for example with only static memory allocation, the use of simple loops with a fixed number of iterations, and a limited set of library functions.

The majority of the restructioning of the code was required to reach a compromise between achieving high computational efficiency by parallel execution and the spending of FPGA resources (chip area). Generally, serial execution of program loops in FPGA is not very efficient, because the clock frequencies are an order of magnitude lower than in standard processors. Faster execution is achieved by parallel execution of loop iterations ("unrolling") and by pipelining of the operations. HLS tools examine the loop structure of the code and include certain automatic optimization of loop execution, but the default optimization settings mostly aim at minimizing spent chip area, so different preferences must be set manually using compiler directives (e.g. `#pragma HLS UNROLL` for parallel execution a loop, `#pragma HLS PIPELINE` for pipelining a loop). Additionally, one must pay attention to the access of variables stored in RAM, because access to RAM is limited, and attempting to read several values from a RAM block at once may prevent unrolling/pipelining of a loop. To work around the RAM access bottleneck, one can split vector or matrix variables into several RAM blocks using the compiler directive `#pragma HLS ARRAY PARTITION`.

Due to the iterative nature of of Algorithm 1, its main iteration loop cannot be parallelized or pipelined. Within the iterations, the matrix-to-vector multiplication $\mathbf{H}_{cp}\mathbf{v}^i$ in the gradient step is the main computational workload. It consists of a two-level for-loop that needs to be optimized. Some details of code adaptation for this multiplication are shown in Appendix 1. To achieve very low computational latency at acceptable use of FPGA resources we applied parallel multiplication of the elements of each row of the matrix `H` with the elements of the vector `v`, followed by tree-wise summation of the resulting products. The outer loop named `L_iter_H_x_v` with the index `ix` spanning the rows of the matrix `H` is pipelined, which implies automatic full parallelization (unrolling) of the inner loops. Within it, the loop named `L_vvmult` with the index `jx` spanning the columns of `H` and the elements of `v` performs the multiplications, and the loops named `L_sum1m1` through `L_sum6m1` (including the final summation to the temporary variable `tvn`) the stages of the binary summation tree, respectively. To facilitate the parallel access to RAM, the compiler directive `#pragma HLS ARRAY PARTITION` is applied to split the rows of `H`, the vector `v` and the summation tree temporary vectors.

The (signed) "ap_fixed" types are used for the FWL representations. Due to the properties of the DSP48E2 digital signal processing blocks available in the target platform, the base bit-width 27 is applied for most variables. The elements of `v` and `H` require 2 and (–1) integer bits, respectively. Hence, products of their elements require 1 integer bit, while their bit-width is restricted to 35 (so that that the bit-width of the sum result `tvn` 27 is reached without loss of accuracy). The code example shows how the bit-widths and numbers of integer bits may be propagated for the intermediate variables of the summation tree (`vt128m` to `vt2m`). However, because the spectral radius of $\mathbf{H}_{cp}$ is 1, the value range of the final stage `tvn` may be assigned the same as that of `v` [20, 29]. For comparison, two floating-point versions of the algorithm were also prepared, where the ap_fixed types are replaced with double or float, respectively.

Finally, the FGM algorithm is equipped with an appropriate communication interface (AXI4) and packaged as an RTL "IP core" block, accompanied with a test-bench function. For actual implementation with the target Xilinx Kintex KCU116 board, the block would be used within a Xilinx Vivado block scheme, also comprising a micro-processor and a PCIe bus interface for the connection with a computer running the tokamak control system. The details are omitted because the hardware implementation was carried out in a different environment.

## 6. Performance evaluation

Due to the hardware availability, the developed FGM algorithm block was tested with a Xilinx Alveo U250 card. Xilinx Alveo are FPGA accelerator card designed for software application acceleration in data-centers. They are programmed using the Eclipse-based Xilinx Vitis 2019.2/2020.1 environment, currently supported only on selected Linux platforms (in particular, we used Ubuntu 18.04 LTS and CentOS 8.1.1911). The Vitis environment includes the HLS block design functionality of Vivado HLS. In addition, it facilitates integrated design and simulation of computational applications on host servers that use FPGA kernels for acceleration. This includes automatically implementing the infrastructure for the communication between the server host application (using the OpenCL API) and the FPGA kernels (which connect to a FPGA shell), either in the form of "AXI4-m" mapping of specific data buffers from the host RAM via the PCIe bus to the FPGA DDR RAM and vice-versa or using "AXI4-stream" streaming communication. We selected AXI4-stream, because with AXI4-m the task enqueue action was observed to cause significant latencies in the range of 0.1 ms.

For use as an accelerator kernel in the Vitis environment, the FGM algorithm block named "RWM_ap" requires a specific kernel header. The input vector bt_d and the output vector u_opt_d are declared as double-valued pointers. Additionally, the communication protocol for them is set to AXI4-stream (axis) with the compiler directive `#pragma HLS INTERFACE`. Additional control signals are transferred using the AXI4lite-slave protocol.

The host application containing the test-bench code is prepared in C++. Instead of simply calling the block function as previously with Vivado HLS, the buffer transfers and kernel execution are conducted using OpenCL API calls as described in Xilinx Vitis documentation [33], chapters 6 and 9. Firstly, `xcl::get_xil_devices` is used to get a list of Xilinx devices in the host computer, and the FPGA binary file is located using `xcl::read_binary_file`. The host application then processes the list of devices, attempting to create the OpenCL context using `cl::Context`, the command queue with `cl::CommandQueue`, and program with `cl::Program`. If the Xilinx card is found and programmed, the host application continues with the creation of a kernel object using `cl::Kernel RWMap_kernel`, the input stream `h2k_stream` and the output stream `k2h_stream` using `xcl::Stream::createStream`, and starts the kernel using `enqueueTask`. In each iteration of the kernel test execution loop, the start time measurement is carried out using `clock_gettime`, a new input vector `x_0` is written to the input `h2k_stream` using `clWriteStream`, the reading of the output `k2h_stream` to the vector `u_opt` is scheduled with `clReadStream`, and the end time is recorded using `clock_gettime`, and finally the result is stored for the assessment of the computation accuracy. When all test data is processed, the OpenCL queue is closed using `q.finish()`, and the streams are released using `xcl::Stream::releaseStream`.

## 6.1. Evaluation results

Table 1 shows the FPGA resource consumption and maximum latency for three variants of the RWM MPC kernel: FWL with 27-bit or higher bit-width, double-precision floating-point and single-precision floating-point. With all three, the input and output vectors were in the double-precision floating-point form, and the conversion was performed internally if required. The results show that the FWL implementation has roughly half the latency and a third of the resource utilization compared to the double-precision floating-point version. The single-precision floating-point version does not consume much more resources than the FWL implementation, but the latency is close to the one of the double-precision version.

Table 1. RWM MPC kernel FPGA resource consumption and maximum latency

| | FPGA resource consumption | | | | Maximum kernel latency | |
|---|---|---|---|---|---|---|
| Kernel variant | BRAM | DSP | Register | LUT | Clock periods | Time (ms) |
| FWL | 83 (3.09%) | 297 (2.42%) | 29681 (0.91%) | 31028 (1.8%) | 3136 | 0.01045 |
| Double-precision floating-point | 168 (6.25%) | 891 (7.25%) | 121958 (3.76%) | 92005 (5.32%) | 6004 | 0.01934 |
| Single-precision floating-point | 83 (3.09%) | 405 (3.30%) | 55498 (1.71%) | 37883 (2.19%) | 5461 | 0.01820 |

Figure 5 shows a waveform diagram of the FWL version of the RWM MPC kernel execution in "hardware emulation". The diagram confirms that kernel execution time for a sample including 20 iterations of the primal FGM algorithm is under 0.011 ms, and that due to the small amount of data the burst read and write transfers take a very small amount of time in comparison. However, the hardware emulation does not include a realistic model of the operation of the PCIe bus, and the actual input/output data transfers may take significantly more time.

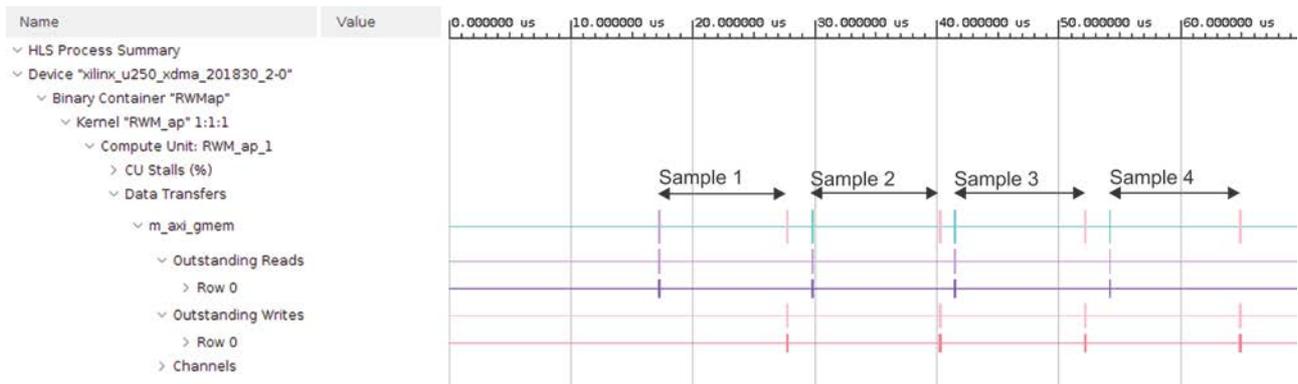

Figure 5. Waveform diagram of hardware emulation of the FWL RWM MPC kernel with AXI4-m communication. The data transfer timelines of the FPGA kernel "RWM_ap" comprising the RWM mpc controller executing a hardware emulation of 4 consecutive samples on the Xilinx Alveo U250 FPGA accelerator board are displayed. "m_axi_gmem" (cyan-violet-red): all memory transfers of the RWM_ap kernel merged; "Outstanding Reads" (violet): short read burst of the state vector **x** to the kernel at the start of each sample; "Outstanding Writes" (red): short write bursts of the computed control vector **u** from the kernel (reads/writes as seen from the FPGA kernel point of view).

Table 2 reports the total (kernel computation and communication) latency measurements conducted using the `clock_gettime` function from the RWM MPC test-bench program with 20 iterations of the primal FGM algorithm running on the host server using the AXI4-stream interface and the worst-case numerical accuracy in terms of $MSE(\tilde{\mathbf{u}}^{20}, \tilde{\mathbf{u}}^{20,*})$ and $J(\tilde{\mathbf{u}}^{20}) - J(\tilde{\mathbf{u}}^{20,*})$. The measurements were carried out with a Xilinx Alveo U250 FPGA accelerator card in a server based on the AMD Ryzen 2950X CPU, running CentOS Linux release 8.1.1911. The best total latencies achieved with the FPGA FWL implementation are: maximum 0.033 ms, on average 0.024 ms. Considering the kernel computation time of 0.011 ms, the average time needed for the communication is 0.013 ms. The maximum values are notably higher that the average ones because these latencies are relatively short and the server was not optimized for low-latency operation. In addition to the three variants of the kernel, previous results obtained using a standard CPU using double- and single-precision floating-point arithmetic [26] are included for reference (the double-precision result serves as the reference solution). The numerical accuracy is considered sufficient both for the FWL and the single-precision floating-point implementation.

Table 2. Total latency measurements and numerical accuracy

| Implementation | Max. latency (ms) | Av. latency (ms) | Max. $MSE(\tilde{\mathbf{u}}^{20}, \tilde{\mathbf{u}}^{20,*})$ | Max. $J(\tilde{\mathbf{u}}^{20}) - J(\tilde{\mathbf{u}}^{20,*})$ |
|---|---|---|---|---|
| FPGA FWL | 0.033 | 0.024 | 5.23·10$^{-5}$ | 1.17·10$^{-6}$ |
| FPGA double-precision floating-point | 0.048 | 0.032 | 1.04·10$^{-13}$ | 7.28·10$^{-11}$ |
| FPGA single-precision floating-point | 0.039 | 0.031 | 3.65·10$^{-5}$ | 6.15·10$^{-7}$ |
| CPU double-precision floating-point | 0.105 | 0.082 | 0 (reference) | 0 (reference) |
| CPU single-precision floating-point | 0.080 | 0.065 | 7.6495·10$^{-5}$ | 8.0717·10$^{-4}$ |

## 7. Conclusions

The MPC controller for RWM control in ITER based on the primal FGM QP solver was implemented in an FPGA using the Xilinx Vivado HLS / Vitis HLS approach. The algorithm was adapted for FWL arithmetic with assistance of Matlab Fixed-Point Converter. Using the Xilinx Alveo U250 accelerator card, the resulting RWM MPC FPGA kernel is able to compute the control signal for each sample in 0.011 ms, which is significantly faster than 0.080 ms previously achieved using an optimized implementation of the same QP solver for a standard CPU. Since the kernel was originally designed for a less-capable FPGA board, the usage of the FPGA resources is modest. The FWL RWM MPC kernel has roughly half the latency of both the single- and double-precision floating-point versions (0.18 ms and 0.19 ms, respectively). The FPGA resource utilization of the FWL kernel is about one third that of the double-precision floating-point variant, while that of the single-precision floating-point version does not increase the utilization considerably.

When the FPGA kernel is used from a host computer, the communication between the host server and the FPGA kernel must also be considered. Initially when using the suggested OpenCL API calls for AXI4-m communication it was observed that the input and output vector transfers took about 0.03 ms each, and the task enqueue action overhead may be close to 0.1 ms. The communication latency was significantly reduced by using AXI4-stream communication, so that the kernel computation task does not need to be enqueued for each sample computation. With this approach, the total MPC controller computation time 33 μs was achieved, measured from the host application. Alternatively, the communication latency can be avoided altogether if control is carried out directly from an FPGA board.

Due to the relatively high demanded solution accuracy in terms of the MSE and the cost function deviation, practically the same control performance of the RWM MPC control scheme as in the original full-precision implementation [26] is expected, despite the limited accuracy of the FWL FPGA computation.

## Acknowledgement

Funding by Slovenian Research Agency (P2-0001) is gratefully acknowledged.

We are grateful to Dept. of Computer Systems, Jožef Stefan Insitute, for letting us use their server with the Xilinx Alveo U250 accelerator card, in particular Anton Biasizzo and Miloš Ljubotina who helped us run the experiments.

Appendix 1: Code adaptation for fast matrix-vector multiplication

This appendix contains snippets of Xilinx HLS C code for FPGA used for the fast execution of the multiplication of the matrix `H` with the vector `v`.

```
#define n_opt_var 81
#define bwidth 27

typedef ap_fixed<bwidth, -1, AP_RND, AP_SAT> apf_H_t;
apf_H_t H[n_opt_var][n_opt_var];
typedef ap_fixed<bwidth, 2, AP_RND, AP_SAT> apf_v_t;
apf_v_t v[n_opt_var];
typedef ap_fixed<bwidth, 2, AP_RND, AP_SAT> apf_tvn_t;
apf_tvn_t tvn[n_opt_var];
typedef ap_fixed<bwidth, apf_H_t::iwidth+apf_v_t::iwidth, AP_RND, AP_SAT> vt128m_t;
vt128m_t vt128m[n_opt_var];

ap_fixed<vt128m_t::width+1+7, vt128m_t::iwidth+1, AP_RND, AP_SAT> vt64m[last64];
ap_fixed<vt128m_t::width+2+7, vt128m_t::iwidth+2, AP_RND, AP_SAT> vt32m[last32];
ap_fixed<vt128m_t::width+3+7, vt128m_t::iwidth+3, AP_RND, AP_SAT> vt16m[last16];
ap_fixed<vt128m_t::width+4+7, vt128m_t::iwidth+4, AP_RND, AP_SAT> vt8m[last8];
ap_fixed<vt128m_t::width+5+7, vt128m_t::iwidth+5, AP_RND, AP_SAT> vt4m[last4];
ap_fixed<vt128m_t::width+6+7, vt128m_t::iwidth+6, AP_RND, AP_SAT> vt2m[last2];

#pragma HLS ARRAY_PARTITION variable=H complete dim=2
#pragma HLS ARRAY_PARTITION variable=v complete
#pragma HLS ARRAY_PARTITION variable=vt128m complete
#pragma HLS ARRAY_PARTITION variable=vt64m complete
#pragma HLS ARRAY_PARTITION variable=vt32m complete
#pragma HLS ARRAY_PARTITION variable=vt16m complete
#pragma HLS ARRAY_PARTITION variable=vt8m complete
#pragma HLS ARRAY_PARTITION variable=vt4m complete
#pragma HLS ARRAY_PARTITION variable=vt2m complete

(...)
        // Tree-wise matrix-to-vector multiplication
        L_iter_H_x_v: for(unsigned char ix = 0; ix < n_opt_var; ix++)
        {
                #pragma HLS PIPELINE    // (unrolls inner loops automatically)
                L_vvmult: for (jx = 0; jx < n_opt_var; jx++) {
                        vt128m[jx] = H[ix][jx] * v[jx];
                }
                vt64m[last64-1] = vt128m[last64-1];
                L_sum1m1: for (jx = 0; jx < imax64; jx++)    {
                        vt64m[jx] = vt128m[jx] + vt128m[jx+last64];
                }
                vt32m[last32-1] = vt64m[last32-1];
                L_sum2m1: for (jx = 0; jx < imax32; jx++)    {
                        vt32m[jx] = vt64m[jx] + vt64m[jx+last32];
                }
                vt16m[last16-1] = vt32m[last16-1];
                L_sum3m1: for (jx=0; jx < imax16; jx++)   {
                        vt16m[jx] = vt32m[jx] + vt32m[jx+last16];
                }
                vt8m[last8-1] = vt16m[last8-1];
                L_sum4m1: for (jx=0; jx < imax8; jx++)       {
                        vt8m[jx] = vt16m[jx] + vt16m[jx+last8];
                }
                vt4m[last4-1] = vt8m[last4-1];
                L_sum5m1: for (jx=0; jx < imax4; jx++)      {
                        vt4m[jx] = vt8m[jx] + vt8m[jx+last4];
                }
                vt2m[last2-1] = vt4m[last2-1];
                L_sum6m1: for (jx=0; jx < imax2; jx++)      {
                        vt2m[jx] = vt4m[jx] + vt4m[jx+last2];
                }
                tvn[i] = (apf_tvn_t) vt2m[0] + vt2m[0+last1];
        }
```

15.01.2021)